\documentclass[format=sigconf, review=false, anonymous=false, screen, dvipsnames]{lib-acm/acmart}

\setcopyright{acmlicensed} %
\copyrightyear{2018}
\acmYear{2018}
\acmDOI{XXXXXXX.XXXXXXX}

\acmConference[Conference acronym 'XX]{Make sure to enter the correct conference title from your rights confirmation email}{June 03--05, 2018}{Woodstock, NY}
\acmISBN{978-1-4503-XXXX-X/18/06}

\acmConference[]{}{}{}
\acmYear{}
\copyrightyear{}
\acmPrice{}
\acmDOI{}
\acmISBN{}
\setcopyright{none}

\usepackage{booktabs} %

\usepackage{exii-macros}

\usepackage{booktabs}
\usepackage{array}
\usepackage{bbding}
\usepackage{tikz}

\hidecomments

\makeatletter
\g@addto@macro\normalsize{%
  \setlength\abovedisplayshortskip{-9pt}
  \setlength\belowdisplayshortskip{3pt}
}
\makeatother

\begin{document}

\tolerance=400 

\title[How Writers Feel About Sharing Prompts in Collaborative Text Editors]{Show Me Your Prompts! How Writers Feel About Sharing Prompts in Collaborative Text Editors}

\author{Nikhita Joshi}
\orcid{0000-0001-9493-7926}
\affiliation{%
\institution{Universit\'{e} Paris-Saclay, CNRS, Inria}
\city{Orsay}
\country{France}
}
\email{joshi@lisn.fr}

\author{Yen-Ting Yeh}
\orcid{0000-0003-4491-8043}
\affiliation{%
\institution{University of Saskatchewan}
\city{Saskatoon}
\country{Canada}
}
\email{allen.yeh@usask.ca}

\renewcommand{\shortauthors}{Joshi and Yeh}

\begin{abstract}
Generative AI writing assistants are becoming integrated into collaborative text editors; however, it is unclear how much information about a user's prompting activities should be shared with collaborators.
We explore the effects of different levels of prompt information sharing within collaborative text editors: not sharing anything, sharing a placeholder to indicate AI use; sharing details about how the resulting text was generated; and sharing everything, including how the prompt was formulated, in real-time.
Sixteen participants wrote persuasive essays in pairs using all four techniques. Results suggest a strong preference for techniques that share more information about prompting activities for increased awareness. Our work shows that collaborative text editors should share more information among writers on when, how, and where AI is used.

\end{abstract}

\begin{CCSXML}
<ccs2012>
   <concept>
       <concept_id>10003120.10003121.10011748</concept_id>
       <concept_desc>Human-centered computing~Empirical studies in HCI</concept_desc>
       <concept_significance>500</concept_significance>
       </concept>
   <concept>
       <concept_id>10003120.10003130.10011762</concept_id>
       <concept_desc>Human-centered computing~Empirical studies in collaborative and social computing</concept_desc>
       <concept_significance>500</concept_significance>
       </concept>
   <concept>
       <concept_id>10003120.10003130.10003233.10011765</concept_id>
       <concept_desc>Human-centered computing~Synchronous editors</concept_desc>
       <concept_significance>500</concept_significance>
       </concept>
 </ccs2012>
\end{CCSXML}

\ccsdesc[500]{Human-centered computing~Empirical studies in HCI}
\ccsdesc[500]{Human-centered computing~Empirical studies in collaborative and social computing}
\ccsdesc[500]{Human-centered computing~Synchronous editors}

\keywords{collaborative writing, human-AI interaction}

\begin{teaserfigure}
  \includegraphics[width=\textwidth]{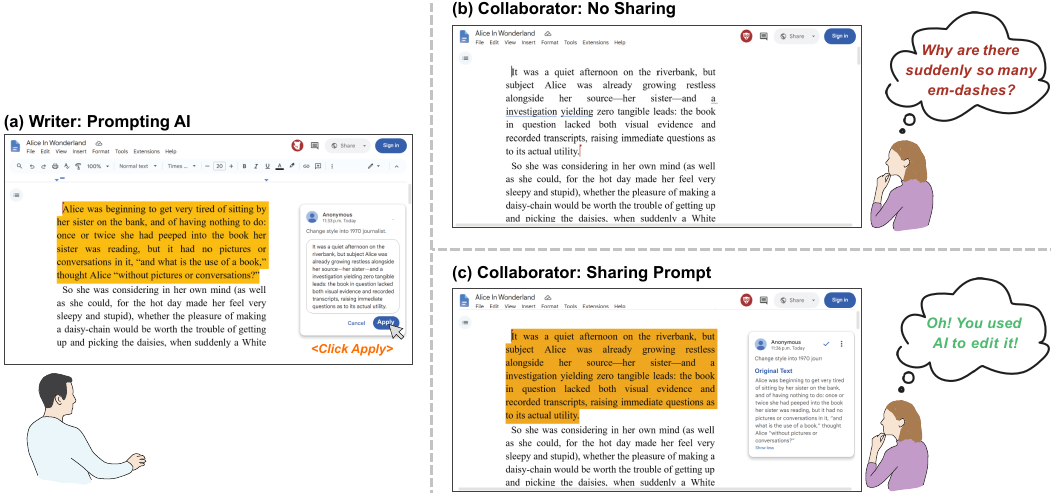}
  \caption{Our work explores different levels of prompt information sharing within collaborative text editors. For example, (a) if someone writes a prompt to generate text, other collaborators could see (b) nothing, or (c) details of how the text was generated.}
  \label{fig:teaser}
\end{teaserfigure}

\maketitle

\section{Introduction}
Different writing activities, such as cursor movements, text selections, and text edits, are immediately shared with others in collaborative text editors like Google Docs and Overleaf. This can improve awareness and collaboration \cite{Dourish1992Awareness}, but it can also distract writers and make them feel self-conscious \cite{Wang2017PrivacyWriting, LarsenLedet2019Territorial, LarsenLedet2020CollabWritingMultipleArtifacts, LarsenLedet2021Idosyncratic}. 

Many people use large language models (LLMs) when writing, for example, \rev{to brainstorm, change the tone of an e-mail, or proofread a paper}. This is typically done by writing prompts in isolation using systems like ChatGPT and pasting the output into a text editor. More recently, LLMs have become integrated into collaborative text editors \cite{GoogleDocGemini, OverleafWritefull, Lehmann2026CollabDocMultipleAgents}, allowing writers to prompt, and insert generated text into a shared document without changing interfaces. 
However, these systems vary greatly in how much prompting activity is shared with collaborators. Despite being collaborative, text editors like Google Docs and Overleaf do not share details of a writer's prompting activities, such as the prompt that was issued and where it was used \cite{GoogleDocGemini, OverleafWritefull}. Alternatively, Lehmann et al.'s text editor \cite{Lehmann2026CollabDocMultipleAgents} integrates this information into comments that are visible for all collaborators to see.

Prompt sharing can benefit other tasks, such as coding \cite{Feng2024CoPrompt}, or prompt engineering, \cite{Reza2025PromptHive}, but it is unclear how much prompting information should be shared with collaborators, and how writers feel about it. \rev{
Our goal is not arguing the need for LLMs when writing, but to address the gap of how to balance group awareness and individual needs when using it.}
Based on prior work on collaborative writing \cite{Yeh2024CoEditing}, we suspect that keeping prompting activities private may help writers feel more comfortable at the expense of collaborator awareness. But this is important to verify, as writing a prompt is not the same as writing text from scratch \cite{Joshi2025Ownership}, \rev{and designers need to know how to create tools that balance group awareness and individual needs.} 

We explore the effects of different levels of prompt information sharing within collaborative text editing tools (\autoref{fig:teaser}). Building on Lehmann et al.'s commenting approach \cite{Lehmann2026CollabDocMultipleAgents}, we designed four techniques that integrate prompting into comments and share different levels of information with collaborators: nothing; a placeholder; details about the prompt that was issued; and all prompting activities, including prompt formulation, in real-time. Together, these techniques explore a design space of sharing \emph{when} AI is actively being used, \emph{how} a response was generated, and \emph{where} AI was used within the document.

We conducted a user study where 16 participants worked in pairs to brainstorm and write persuasive essays using all techniques. Participants strongly preferred techniques that share more information because of the benefits associated with increased awareness:
\rev{not leaving each other `hanging,' following each other's thoughts, improving trust and learning, and the ability to verify AI-produced text.}
Our work contributes empirical results that motivate more prompt information sharing within collaborative text editors.

\section{Background and Related Work}

Our work relates to prior work on prompt sharing, multi-user collaborative writing with AI, and how users feel about writing together.

\subsection{Prompt Sharing when Collaborating}
Prior work has shown that prompt sharing can be beneficial for a variety of tasks.
Feng et al.'s CoPrompt system \cite{Feng2024CoPrompt} allows collaborators to share prompts with each other when writing code for data analysis. Users can view prompt histories, create links between prompts, and request information from other prompts, which helped users better understand each other's workflows.
Recognizing that writing effective prompts can be difficult \cite{JD2023WhyJohnny}, Reza et al.'s PromptHive system \cite{Reza2025PromptHive} allows educators to generate hints for classroom assignments by prompting an LLM. Prompts are posted to a public ``prompt library,'' where users can view and build on each other's prompts. This helped users craft more effective and creative prompts.
Though not focused on prompt sharing, Han et al. \cite{Han2024CoWritingPrompts} had pairs co-write prompts to generate images, and found that it encouraged more discussion and helped develop a mutual understanding of each other's ideas. 

Researchers have identified clear advantages of prompt sharing and increased awareness of a collaborator's prompting activities. However, little is known about the impact of different levels of prompt information sharing when using AI, yet alone while writing collaboratively.

\subsection{Multi-User Collaborative Writing with AI}
Several systems have been designed to support human-AI collaborative writing (cf. Lee et al.'s literature review \cite{Lee2024DesignSpaceWriting}), but relatively few involve multiple users. 
Google Docs \cite{GoogleDocGemini} and Overleaf \cite{OverleafWritefull} allow writers to prompt LLMs in a standalone chat. Users can reference their document by selecting text before submitting their prompt, and then decide whether to accept the AI-generated text.
This is kept private from other collaborators: they simply see the new text appear, as if it was pasted in. 

More recently, Lehmann et al. \cite{Lehmann2026CollabDocMultipleAgents} integrated prompting into document comments that are visible for all collaborators to see. Writers can select text, type a prompt, and view the AI-generated text within a comment anchored to that selection. Changes are integrated into the document after accepting them. Since choosing to accept a comment became a team decision, participants felt like they collaborated more effectively.

Even from these few examples, we notice stark differences in how much prompting activity is shared with collaborators. As generative AI tools become more integrated into collaborative text editors, we believe it is important to understand how much information should be shared with collaborators and their pros and cons.

\subsection{Collaborative Writing Sentiments}
Prior work suggests that people often feel uncomfortable when writing collaboratively. Wang et al.'s interviews \cite{Wang2017PrivacyWriting} revealed that many writers do not want to write together, even when they are working on collaborative writing tasks. This has been corroborated by others, such as Larsen-Ledet and colleagues \cite{LarsenLedet2019Territorial, LarsenLedet2021Idosyncratic, LarsenLedet2020CollabWritingMultipleArtifacts}. Specifically, writers become distracted by their collaborator's activities, including live edits and changes in cursor position, and express concern over being judged when their own writing activities are being watched.

Writers have adopted strategies to feel more comfortable, such as indicating personal writing regions within shared documents \cite{LarsenLedet2019Territorial}, and temporarily writing in a private document and pasting the contents back into the shared document \cite{LarsenLedet2019Territorial, Wang2017PrivacyWriting, Strobl2014ForeignWriters}. Researchers have also explored ways of giving writers more control over when and how their writing activities are shared with others. Ignat et al. \cite{Ignat2008PrivacyPreserving} created a text editor that allows writers to choose what activities are shared with their collaborators and applies visual filters, such as blur effects, to the edited text to preserve privacy. Yeh et al. \cite{Yeh2024CoEditing} proposed manipulating when and how updates are shared with collaborators, such as only sharing edits with collaborators after an entire sentence is written. Their user studies showed that slower updates that hide more writing helped writers feel more comfortable, but at the expense of collaborator awareness.

Our work follows a similar approach as Yeh et al. as it also explores how writers feel about different levels of information sharing, but focused on prompts rather than document edits. Based on this literature, it is likely that strategies that share less information about prompting activities will help writers feel more comfortable, but reduce collaborator awareness. Still, it is important to confirm that this is the case, as prior work suggests that writers feel differently about text that is written with AI assistance \cite{Joshi2025Ownership}.
As such, writers may feel less self-conscious or fearful over being watched, and may value increased collaborator awareness instead.

\medbreak

To summarize, prior work shows that prompt sharing is beneficial, but it is unclear how much information should be shared when writing collaboratively, a task that is commonly associated with discomfort. Our work builds on these concepts by evaluating how much information writers believe is appropriate to share about their prompts with their collaborators.

\section{Levels of Prompt Information Sharing}

Dourish and Bellotti \cite{Dourish1992Awareness} note that sharing more information with collaborators can improve \emph{awareness}: ``understanding of the activities of others, which provides a context for your own activity.''
This can be achieved by automatically sharing information about a user's actions, such as what they are working on and how their actions relate to the task. Awareness ``is critical to successful collaboration,'' \cite{Dourish1992Awareness} and is therefore important to have in collaborative tools.

The generative AI tools in Google Docs and Overleaf \cite{OverleafWritefull, GoogleDocGemini} do not follow this. When a user begins writing a prompt, other users have no indication of this action, so they do not know \emph{when} AI is being used. After the user accepts a suggested response, AI-generated text is integrated into the document without any information on \emph{how} the text was generated as the prompt is not shared. This could make understanding its relevance to the collaborative writing task more difficult. Related, this AI-written text is not distinguished from human-written text, so other users have no way of knowing \emph{where} AI was used.
Based on these issues, we explore a design space of prompt information sharing across three dimensions:

\begin{figure*}[p]
    \centering
    \includegraphics[width=\textwidth]{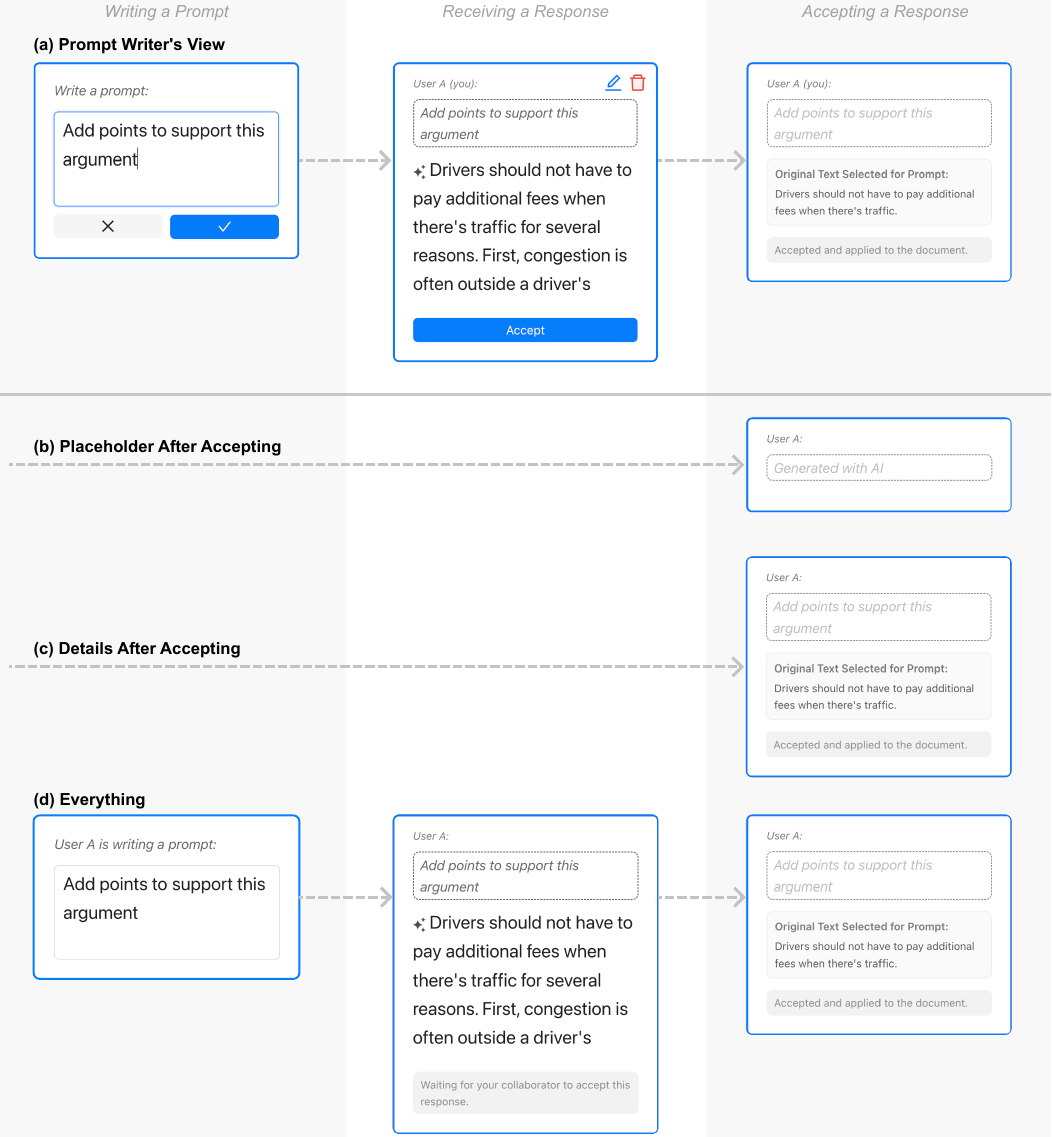}
    \caption{What (a) prompt writers and other collaborators see when they see (b) a placeholder after the response is accepted; (c) details after the response is accepted; and (d) everything. Other collaborators cannot see anything when nothing is shared.}
    \label{fig:techniques}
\end{figure*}

\begin{enumerate}
    \item \textbf{\emph{when}} AI is being used by indicating when the user is actively typing a prompt;
    \item \textbf{\emph{how}} AI is being used by revealing information that was used to generate the resulting text, such as the prompt that was issued; and
    \item \textbf{\emph{where}} AI is being used by indicating regions in the document that  were generated by AI. 
\end{enumerate}

\begin{table}[t]
    \centering
    \caption{Design space of prompt sharing strategies we explored across three dimensions: \emph{when}, \emph{how}, and \emph{where}.}
    \renewcommand{\arraystretch}{1.3}

\centering
\begin{tikzpicture}
    \node (table) {
        \begin{tabular}{l|c|c|c}
        \toprule
        \textbf{Strategy} & \textbf{\emph{When}} & \textbf{\emph{How}} & \textbf{\emph{Where}}\\
        \midrule
        Nothing &  & & \\\hline
        Placeholder After Accepting & & & \checkmark \\ \hline
        Details After Accepting & & \checkmark & \checkmark \\ \hline
        Everything & \checkmark & \checkmark & \checkmark \\
        \bottomrule
        \end{tabular}
    };

    \draw[->, line width=\heavyrulewidth] 
        ([xshift=-0.1cm]table.north west) --
        ([xshift=-0.1cm]table.south west)
        node[midway, rotate=90, above, yshift=2pt] {Shared Information};
\end{tikzpicture}
    \label{tab:techniques}
\end{table}

To explore this design space, we followed Lehmann et al.'s approach \cite{Lehmann2026CollabDocMultipleAgents} and integrated prompting into comments.
We do this for three primary reasons. First, users prefer this technique when reading \cite{Joshi2026MarginNotes} and writing \cite{Lehmann2026CollabDocMultipleAgents}. Second, leaving comments is a common feature in collaborative text editors. As such, users would likely focus more on the newer concepts around prompt sharing.
Third, the comment's visibility and contents can be easily manipulated in ways that allow us to explore all three design dimensions.

\subsection{Prompting with Comments}
To create a comment, the user first selects text from the document. This causes a ``Prompt'' button to appear beside the text selection. Clicking it opens a text box in the right margin, beside the text editor. Here, the user can write a prompt for an LLM and submit it by pressing a ``Confirm'' button. A few seconds later, the LLM produces a response. 
The user can edit the comment to generate a new response, or delete it. If they want to integrate the resulting text into the document, they can press an ``Accept'' button.

The comment and its associated prompt remain on the screen after the text has been integrated into the document. Clicking the comment highlights the generated text and expands to reveal the original text (\autoref{fig:techniques}a).

\subsection{Prompt Information Shared with Others}
Other collaborators may see different information, providing more or less awareness of \emph{when}, \emph{how}, and \emph{where} AI was used in the document (\autoref{tab:techniques}). \emph{All techniques are shown in the supplementary video.}

\subsubsection{Nothing}
No comment is created. As a result, collaborators are unaware of \emph{when}, \emph{how}, and \emph{where} AI is used. This best reflects the mechanics of commercial tools like Google Docs \cite{GoogleDocGemini} and Overleaf \cite{OverleafWritefull}, and current practices of prompting in a separate tool like ChatGPT and pasting the contents into a shared document.

\subsubsection{Placeholder After Accepting}
After the prompt writer accepts a response and the AI-generated text is merged with the document, collaborators only see a placeholder prompt that says ``Generated by AI'' (\autoref{fig:teaser}b). They do not see the original text selection or the prompt that was used to produce the new text. This gives collaborators awareness of \emph{where} AI was used, but not details of \emph{how} or \emph{when} it was produced.

\subsubsection{Details After Accepting}
Collaborators see the same information as the prompt writer after they accept a comment: the prompt and the original text selection that was used to generate the text (\autoref{fig:teaser}c). This gives collaborators awareness of both \emph{where} and \emph{how} AI was used, but not details of \emph{when} it was used as updates are not shared in real-time.

\subsubsection{Everything}
Collaborators see all phases of the comment creation process in real-time (\autoref{fig:techniques}d). They see their collaborator write a prompt, receive and accept the LLM's response live.
This gives collaborators complete awareness of \emph{where} AI was used, even more information about \emph{how} it was used, and \emph{when} it was used.

\section{User Study}
The goal of this user study is to better understand how writers feel about different levels of prompt information sharing. Participants wrote persuasive essays in pairs by brainstorming together before writing individual paragraphs while their partner observed.

\subsection{Participants}

\begin{table*}[h!]
    \centering
    \caption{Participant demographics for our user study. For Comfort, higher scores correspond to feeling more comfortable when writing collaboratively.}
    \small %
\begin{tabular}{lr|lr|lr|lr|lr|lr}
\toprule
\multicolumn{2}{l|}{Gender} & \multicolumn{2}{l|}{Age} & \multicolumn{2}{l|}{Education} & \multicolumn{2}{l|}{English Writing Proficiency} & \multicolumn{2}{l|}{Relationship Duration} & \multicolumn{2}{l}{Comfort}\\
\midrule
Men & 9 & 18-24 & 5 & Some University (no credit) & 3 & Native or Bilingual & 10 & $<$ 1 Year & 7 & 2/7 & 1\\
Women & 7 & 25-34 & 9 & Bachelor's Degree & 7 & Full Professional & 6 & 1-3 Years & 3 & 5/7 & 7\\
&& 35-44 & 2 & Master's Degree & 4 &&& 3-5 Years & 2 & 6/7 & 3\\
&&&& Professional Degree & 1 &&& $>$ 5 Years & 4 & 7/7 & 5\\
&&&& Doctorate Degree  & 1\\
\bottomrule
\end{tabular}

\begin{tabular}{lr|lr|lr|lr}
\\
\toprule
\multicolumn{2}{l|}{Editor Frequency} & \multicolumn{2}{l|}{Commenting Frequency} &\multicolumn{2}{l|}{LLM Frequency} & \multicolumn{2}{l}{LLM Writing Frequency}\\
\midrule
Daily & 5 & Daily & 2 & Daily & 10 & Daily & 5\\
Weekly & 4 & Weekly & 3 & Weekly & 3 & Weekly & 4\\
Monthly & 3 & Monthly & 3 & Less than Monthly & 1 & Monthly & 2\\
Less than Monthly & 4 & Less than Monthly & 8 & Never & 2 & Never & 5\\

\bottomrule

\end{tabular}
    \label{tab:demographics}
\end{table*}

We recruited 8 pairs (16 participants; \autoref{tab:demographics}). Participants were recruited using university mailing lists and word-of-mouth, and received \$20 upon successful completion of the study. To better represent cases where collaborative writing would occur, both participants in each pair knew each other prior to participating.

\subsection{Apparatus}

\begin{figure*}[t]
    \centering
    \includegraphics[width=\textwidth]{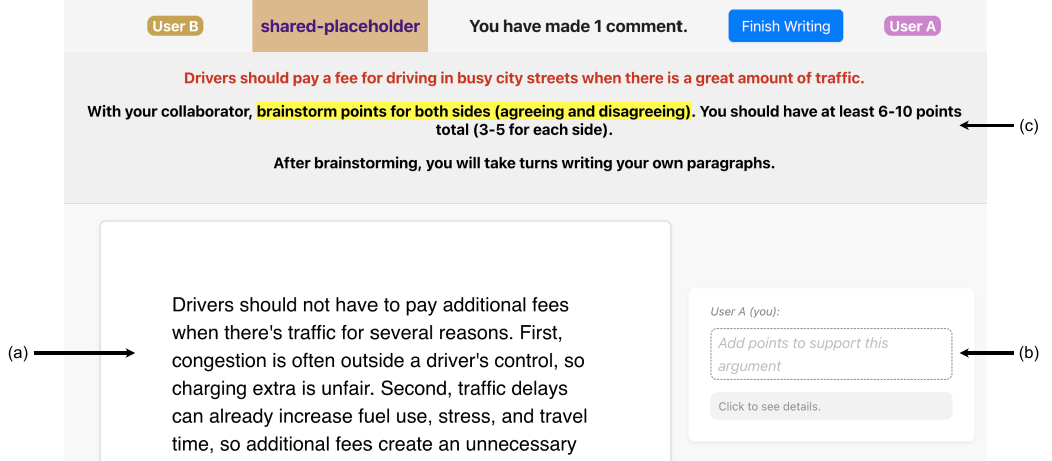}
    \caption{The main interface used in the user study: (a) a collaborative text editor, (b) comments, and (c) a toolbar containing instructions.}
    \label{fig:apparatus}
\end{figure*}

We created a custom web application using Node.js and React. A collaborative text editor was displayed on the left, giving participants a space to write (\autoref{fig:apparatus}a). This was implemented using a Quill text editor. 
Comments appeared to the right of the text editor (\autoref{fig:apparatus}b), and all responses were generated using GPT-5.4 Mini.
All changes to the document text and the comments were synchronized using Yjs and a custom Hocuspocus web server. At the top was a header that contained instructions and a ``Finish Writing'' button that participants pressed to end the trial (\autoref{fig:apparatus}c).

\subsection{Task}
Participants worked in pairs to complete persuasive writing tasks, which were based on statements from Stab et al. \cite{Stab2017Essays} (e.g., \emph{``Learning about the past has no value for those of us living the present''}). First, participants worked together for 5 minutes to brainstorm three to five arguments that agreed and disagreed with the provided statement.

Next, each participant was assigned a random stance to support. One participant wrote a paragraph for their stance for approximately four minutes, using their brainstormed ideas. Meanwhile, the other participant observed them. This setup has shown to be effective in prior work \cite{Yeh2024CoEditing} for improving understanding of different conditions and mimicking the uncomfortable writing scenario of being watched. After, participants switched roles and repeated the process. 

To ensure that participants had some experience prompting with and observing the different prompt sharing strategies, they had to have at least three accepted comments for each stage. This number was refined during pilot tests and has been used in prior work \cite{Joshi2026MarginNotes}.

\rev{We believe this task approximates one that would be experienced in an educational context (writing an essay together). As per Baecker et al.'s collaborative writing taxonomy \cite{Baecker1993WritingRoles}, it also features many properties that are common among collaborative writing tasks: two activities (``brainstorming'' and ``writing'') and two strategies (``joint writing'' and ``separate writing'').}

\subsection{Procedure}
The study took place online. After obtaining informed consent 
and providing demographic information, participants listened to a short presentation that provided general instructions for the entire study. Next, they read instructions about the prompt sharing technique, and tried it in a tutorial document for approximately 3 minutes.

They then began the specific trial, switching between brainstorming, writing, and watching. After that, they completed a short questionnaire about their experience with the technique. This repeated for all four techniques. Finally, they answered a final questionnaire about their overall experience. The entire study took approximately 100 minutes.

\subsection{Design}

This is a within subjects design with one primary independent variable: 
\f{condition} with 4 levels (\f{nothing}, \f{placeholder}, \f{details}, \f{everything}). The order of \f{condition} was counterbalanced using a Latin square. Participants received a random topic to write about from Stab et al. \cite{Stab2017Essays}, and were assigned a random stance (agree or disagree) for the writing stage. The order of watching or writing was switched between conditions. 

\subsection{Data}
We collected free-form responses and obtained the following metrics from questionnaires (all 1-7 unless indicated otherwise), which span four categories.

\subsubsection{Writer Experience}
This represents how participants felt about writing text using each technique. This was done by asking three Likert questions about \m{Writer Comfort}, how comfortable they felt; \m{Writer Information Satisfaction}, how satisfied they felt about the information that was shared; and \m{Writer Frequency Satisfaction}, how satisfied they felt about the frequency of shared updates.

\subsubsection{Observer Experience}
This represents how participants felt when they saw their partner use each technique. This was also done by asking three Likert questions about \m{Line of Thought}, how much they could understand their partner's thinking; \m{Observer Information Satisfaction}; and \m{Observer Frequency Satisfaction}.

\subsubsection{Workload}
We collected information about \m{Mental Demand}, \m{Effort}, and \m{Frustration} using questions from the NASA-TLX \cite{NASATLX}. Note that we did not ask about physical demand, temporal demand, or performance, as prior work suggests these are not relevant factors for this type of task \cite{Joshi2026MarginNotes}.

\subsubsection{Preferences}
We asked about \m{Frequency of Use}, how frequently the participant would use the technique if made available to them, which was taken from the SUS \cite{brooke1996sus}. 
After trying all conditions, participants ranked the four conditions in order of preference to provide an \m{Individual Ranking} (1-4). Ties were allowed. To establish an \m{Overall Ranking} (1-4), the \m{Individual Ranking} was analyzed using the Condorcet method \cite{young1988condorcet}. The \f{condition} ranked 1st outperforms all others in pairwise comparisons, the one ranked 2nd outperforms all save the one ranked 1st, and so on.

\section{Results}

Where applicable, we use Friedman omnibus tests and Wilcoxon signed-rank post hoc tests with Holm's corrections for multiple comparisons, and Spearman's correlations. Free-form responses were coded and grouped into themes by the first author following an inductive thematic analysis approach \cite{Braun2006ThematicAnalysis}. 
The codes and themes were iteratively refined through discussions with the second author.

We did not observe any meaningful significant differences for \m{Writer Comfort}, \m{Frequency of Use} or any workload-related factors, so we focus our analysis on other metrics.

\subsection{Sharing More Information is Preferred}
\begin{figure}[t]
    \centering
    \includegraphics[width=0.47\textwidth]{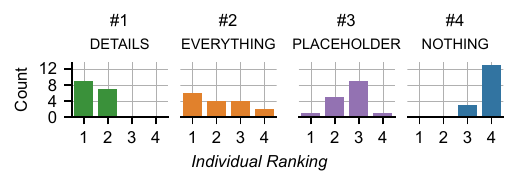}
    \caption{Distribution of \m{Individual Ranking} counts. The order of \f{condition} corresponds to the \m{Overall Ranking}.}
    \label{fig:rankings}
\end{figure}

Our results suggest that participants preferred techniques that share more information. \f{details} had an \m{Overall Ranking} of 1, followed by \f{everything}, \f{placeholder}, and \f{nothing}. Notably, all participants assigned a first- or second-place \m{Individual Ranking} to \f{details}, and most (62.5\%) did so for \f{everything}. In contrast, the majority of participants (62.5\%) gave a third- or fourth-place \m{Individual Ranking} to \f{placeholder}, and all did so for \f{nothing} (\autoref{fig:rankings}).

\begin{figure*}[t!]
    \centering
    \includegraphics[width=\textwidth]{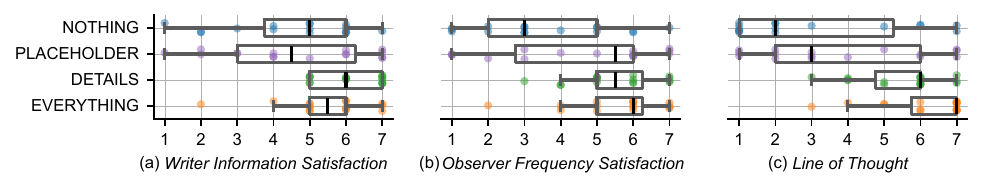}
    \caption{Responses for (a) \m{Writer Information Satisfaction}; (b) \m{Observer Frequency Satisfaction}; and (c) \m{Line of Thought}.}
    \label{fig:metrics}
\end{figure*}

This preference for increased sharing was supported by other metrics. There was a significant effect of \f{condition} on \m{Writer Information Satisfaction} (\friedmanNEffect{3}{16}{10.43}{.01}{.22}; \autoref{fig:metrics}a), with post hoc tests showing that participants were more satisfied with \f{details} (\median{6}, \iqr{2}) than \f{nothing} (\median{5}, \iqr{2.25}; \p{.05}). 
Similarly, there was a significant effect of \f{condition} on \m{Observer Frequency Satisfaction} (\friedmanNEffect{3}{16}{11.01}{.01}{.23}; \autoref{fig:metrics}b). Post hoc tests showed that participants were more satisfied with \f{details} (\median{5.5}, \iqr{1.25}) and \f{everything} (\median{6}, \iqr{1.25}) than \f{nothing} (\median{3}, \iqr{3}; both \p{.05}).

Free-form responses suggest that all participants valued having increased awareness, for example: \ppquote{I prefer having my collaborator see what I'm prompting the AI with, and for them to see what I'm doing}{P09}.
We identified five core benefits, which we grouped based on the three dimensions of our design space.

\subsubsection{Knowing When AI is Used}
Six participants (37.5\%) noted how non-real-time updates made it difficult to tell when prompts were being written. Specifically, \ppquote{it looked like they were doing nothing until they wrote the prompt and the AI finished}{P09}, which \ppquote{leaves [them] in suspense}{P04}. Real-time updates with \f{everything} helped writers \ppquote{feel less nervous about leaving [collaborators] hanging}{P04}.

\subsubsection{Knowing How AI is Used}
Fifteen (94\%) noted how sharing more information made it easier to follow each other's line of thought (or that sharing less information made it more difficult), for example: \ppquote{I think that sharing my prompts with my collaborator can at times make them more aware of the angles that I am going for}{P10} and \ppquote{[with \textsc{nothing}], I had no insight [into] what my collaborator was thinking or trying to accomplish}{P01}.

These sentiments were supported by a significant effect of \f{condition} on \m{Line of Thought} (\friedmanNEffect{3}{16}{19.6}{.001}{.41}; \autoref{fig:metrics}a). Post hoc tests showed that \f{details} (\median{6}, \iqr{1.25}) and \f{everything} (\median{7}, \iqr{1.25}) had significantly higher scores than \f{placeholder} (\median{3}, \iqr{4})  and \f{nothing} (\median{2}, \iqr{4.25}; all \p{.05}).

Participants (4, 25\%) also felt that sharing more information with their collaborators improved trust (and that not sharing information eroded it). For example, participants mentioned how \f{nothing} felt \ppquote{deceptive}{P14} and \ppquote{can cause issues in trust with the collaborator}{P06}, and how \f{placeholder} felt \ppquote{like hiding things}{P11}. Interestingly, one participant noted how \f{placeholder} could cause even more distrust, as it tells collaborators that an LLM generated the text, which warrants closer attention, but necessary information to judge it is not provided: \ppquote{[it] doesn't convey any additional useful information and actually increases distrust in the text}{P08}.

Three participants (19\%) were generally curious to see how their collaborators used LLMs, and five (31\%) believed that sharing more information could help them learn new prompting strategies, for example: \ppquote{it was interesting to see how my collaborator prompted AI. I will probably steal some of their methods}{P10} and \ppquote{[we] can learn better from each other on how to prompt and change things}{P06}. One participant even tried new prompting strategies during the study, and was disappointed when prompt details were hidden as they could not learn: \ppquote{[with \textsc{details}], I saw that my collaborator asked AI to make the paragraph more persuasive, and so I used that in my prompt this time. [With \textsc{nothing}], I did not know what my collaborator asked in the prompt, and so I could not benefit from any of their ideas}{P07}. %

\subsubsection{Knowing Where AI is Used}
Participants (7, 44\%) wanted to easily know where AI was used in the document, for example: \ppquote{if I am collaborating with someone and they are using AI, I'd like to be able to know which part of the sentence they used the AI for}{P06}. A major reason for this is related to contentiousness: five participants (31\%) wanted to double check the accuracy of text that was produced by an LLM, supported by responses like: \ppquote{I'm fine showing them that I used AI, in fact I would prefer that they know, so they can correct me if I am wrong somewhere}{P15} and \ppquote{AI makes a lot of mistakes. It is a necessity for a human [to] proofread when something is written by AI}{P06}.
This was especially challenging to do with \f{nothing}, as no comment was created after a response was accepted, for example: \ppquote{when nothing is shared, it is very tough to understand what approach my collaborator is using. I have to read the whole thing to understand every time they make a change}{P05}.

\subsection{Sharing Comes with Challenges}
Despite sharing the most information, \f{everything} was not ranked first, suggesting that there are challenges that could prevent participants from sharing all prompting activities in real-time. Using insights from free-form responses, we present five reasons that may have contributed to this, organized into prompt writer and observer perspectives. 

\subsubsection{Writer Experience}
Some participants (5, 31\%) felt that prompting was a private activity, for example: \ppquote{sharing my prompts with my collaborator kind of violates [my] writer's privacy}{P16}. Although we did not observe a significant effect of \f{condition} on \m{Writer Comfort}, 
half (8, 50\%) described feeling uncomfortable when sharing their prompting activity. This was especially the case when updates were shared in real-time with \f{everything}. For example, participants disliked \ppquote{being watched}{P04}, noted feeling \ppquote{more self-conscious}{P11}, worried about \ppquote{feeling judged}{P01}, and felt \ppquote{a bit embarrassed}{P07}. 
When updates were not provided in real-time, four participants (25\%) noted how this helped them feel more comfortable: \ppquote{[With \textsc{details}], I was more confident in writing the prompts because I was not scared of messing up. [...] I don't want them to think I am stupid}{P06}. 

Similarly, six participants (37.5\%) expressed specific fear over how their collaborators would interpret the quality of their prompts, which was mitigated by techniques that shared less information. For example, \ppquote{I had more peace of mind knowing my short simple prompts weren't shared with the collaborator}{P06}; \ppquote{if I make sloppy prompts, I would rather not my collaborator see it}{P05}; and \ppquote{I don't have to think too much about the wording of the prompt so it doesn't sound too goofy or lazy}{P11}. However, only one participant admitted to judging the prompts used by their collaborator: \ppquote{I wonder if my collaborator might be judging the quality of my prompts, because I will sometimes find myself judging the quality of his prompts. [...] I had to try not to judge them though, sometimes they seemed too simple or mundane, like `what's the point of even using a prompt for that task?'}{P13}. 

Four participants (25\%) described how prompt sharing caused them to change behaviour. For example, some described adopting better prompting strategies when they knew their prompts were being shared: \ppquote{I knew my prompts were shared with someone else, and naturally I had to focus on proper use of prompts}{P16}. Two felt that seeing their collaborator's prompting activity influenced their own thoughts, for example: \ppquote{[seeing their prompt] hijacked my own thought process and prevented ideas I would have come up with on my own}{P12} and \ppquote{less information gives me more room to think as a writer}{P04}.

\subsubsection{Observer Experience}
Others (4, 25\%) had personal opinions that seeing their prompt formulation in real-time was not important, for example: \ppquote{I don't mind not knowing the exact prompt until my collaborator is done}{P09}. It also prevented them from seeing unnecessary information: \ppquote{if they made a mistake and erased it, it is not worth seeing}{P05}.

Some participants (3, 19\%) felt that viewing more information about their collaborator's prompts in real-time with \f{everything} was \ppquote{very annoying}{P14} and \ppquote{pretty distracting}{P01}, and may have affected their ability to write, for example: \ppquote{it was more distracting [and took] from my ability to focus on writing}{P14}. These suggest some frustration; however, we did not observe a significant effect of \f{condition} on \f{Frustration}.

\subsection{Comparing Writer and Observer Roles}
Taken together, these remarks about the perceived benefits and challenges of sharing could explain why some techniques are perceived better as a prompt writer than as an observer. 
Considering \m{Information Satisfaction} and \m{Frequency Satisfaction}, we found that participants tended to be more satisfied as a writer for \f{placeholder} (both \p{.05}). In contrast, for \f{everything}, participants were significantly less satisfied with \m{Update Frequency} when they were a writer (\p{.05}). 

This was corroborated by free-form responses. Half noted trade offs and differences in preferences based on their role, for example: \ppquote{more private measures would allow for greater anonymity in terms of your workflow but it would make it more confusing for the observer}{P14}; and \ppquote{I get a kick out of seeing the real time thought process, but not if they can see mine}{P12}.

\section{Discussion}

In summary, our results suggest that participants preferred techniques that shared more prompting activities. Participants noted multiple perceived benefits related to increased awareness that came from sharing more information. Specifically, they wanted to know \emph{when}, \emph{how}, and \emph{where} AI was being used.

Participants could more easily tell what their collaborators were doing when they knew \emph{when} prompts were being formed in real-time. Sharing the prompt and original text selection gave insights into \emph{how} the resulting text was formed. This helped participants understand each other's thought process more, built trust, and provided opportunities to learn from each other. Leaving a comment signalled \emph{where} AI was used, which was important as participants wanted to indicate and know parts of the document that needed additional verification.

That said, there were some perceived cons to sharing more information about prompting activities. Some participants noted that formulating a prompt should be ``private,'' felt uncomfortable when everything was shared, and worried the quality of their prompts would be judged. Having awareness of other's prompts may have influenced each other's thinking in ways they did not appreciate and found distracting and annoying.

We discuss design implications that come from these results, how our results compare to prior work, other designs that could be considered, limitations, and other possibilities for future work.

\subsection{Design Implications}
\label{sec:designImplications}
\emph{Collaborative text editors with integrated AI writing assistants should make prompting details available for others to see}, given the many benefits of increased awareness.

The results suggest that \emph{sharing details, such as the original text selection and prompt, after a comment is accepted might strike the best balance} as writers have more awareness of \emph{how} the resulting text was formed and \emph{where} AI was used, 
without fear of being watched when all details of prompt formulation are shown.

However, this approach does not give enough insight into \emph{when} AI is being used, something that participants valued when everything was shared. As such, another implication for design is to \emph{devise ways of sharing when users are writing prompts that still maintain privacy}. One possibility is to show when users are writing prompts in real-time with a placeholder (e.g., ``Bob is writing a prompt...''). This would give writers indication of \emph{when} AI is being used before the response is accepted, \emph{how} it was used by still sharing the prompt, and \emph{where} it was used by leaving a comment for others to see.

\subsection{How Prompting and Writing Text Differ}
Some of our results align with prior work studying collaborative writing without AI. Like Wang et al. \cite{Wang2017PrivacyWriting}'s concept of ``the privacy of writing,'' we found that many participants also view prompting to be a private activity, suggesting that there is a ``privacy of prompting:'' personal beliefs that prompting should be done privately.
Similar to work by Larsen-Ledet and colleagues \cite{LarsenLedet2019Territorial, LarsenLedet2020CollabWritingMultipleArtifacts, LarsenLedet2021Idosyncratic}, our results indicate some discomfort over being watched when writing prompts, but also distraction from real-time updates. 
Our results also suggest a similar trade off as Yeh et al. \cite{Yeh2024CoEditing}, who highlighted several tensions of non-real-time sharing, such as hiding immediate writing at the expense of collaborator awareness.

However, unlike prior work, \emph{our results suggest that writers are more tolerant of these issues when they are prompting}, as evidenced by the high overall rankings given to techniques that share prompts, even in real-time. 
There may be a few reasons for this. One is that the perceived advantages of sharing outweighed the disadvantages. Certain benefits like building trust, learning prompting styles from each other, and verifying the output, may have been especially important when writing with AI. These benefits were not considered in prior work.
Another possibility is that participants did not feel much psychological ownership for the resulting text \cite{Joshi2025Ownership} and that collaborators knew the resulting text was not truly their collaborator's writing, so they did not care about sharing their prompts. This was supported by free-form responses like: \ppquote{you can see someone else's direction of the thought, but the execution isn't theirs}{P12}, but more work is needed to better understand the relationship between prompt sharing preferences and psychological ownership.

\subsection{Other Designs}

We chose to integrate LLM capabilities into comments like prior work \cite{Joshi2026MarginNotes, Lehmann2026CollabDocMultipleAgents}, but there are many other possibilities, such as Google Docs' approach \cite{GoogleDocGemini} of integrating responses directly into the document. 

Even within the continuum we explored, there are also other designs that could be explored, such as providing feedback of \emph{when} prompts are being typed, but not \emph{where} by not sharing a comment. This fall in between the \f{nothing} and \f{placeholder} conditions. Similarly, sharing details before the response is accepted could encourage even more discussion and collaboration among writers, as everyone could decide whether a response should be accepted \cite{Lehmann2026CollabDocMultipleAgents}. This could lead to even more trust and learning. An even more involved approach is to allow writers to co-write prompts \cite{Han2024CoWritingPrompts}, which could help them have even more awareness and understanding, but would likely be more distracting.

Designing and comparing alternative ways of integrating generative AI into collaborative text editors is an exciting direction for future work, and we believe that our results will help designers think about the ways of sharing \emph{when}, \emph{how}, and \emph{where} AI is used.

\subsection{Limitations and Future Work}
We outline some limitations of our approach and highlight other opportunities for future work.

\subsubsection{Ecological Validity}
Our task required participants to switch between brainstorming and watching each other write individually. This covers the ``brainstorming'' and ``writing'' activities and the ``joint writing'' and ``separate writing'' strategies from Baecker et al.'s collaborative writing taxonomy \cite{Baecker1993WritingRoles}. It also gives writers adequate exposure to form opinions \cite{Yeh2024CoEditing} as both a prompt writer and an observer seeing someone else prompt, but it is not representative of all collaborative writing. 
\rev{Furthermore, the experimental setting may have increased participants' expectations for awareness of their collaborator's activities, as they were given specific tasks and goals to achieve within a limited time. In reality, synchronous collaboration would occur less frequently, and collaborators would have fewer expectations of continuously monitoring each other's activity. A longitudinal study would complement and extend our findings by showing how sharing preferences evolve over time, and across different tasks.}

For increased control, our apparatus was limited to certain features and it is possible that this affected sentiments. For example, participants could not ``undo'' accepting a response, and unlike Google Docs' ``suggestions'', accepting a response did not make the comment disappear. Similarly, a deployed system would likely have other quality of life features, such as the ability to edit a generated response before integrating it into the document, and the ability to iterate over multiple prompts. Having these features may help writers feel even more comfortable about sharing their prompting activities.

\subsubsection{Other Scenarios}
Some participants mentioned that their feelings might change in different collaborative scenarios. We recruited pairs of participants who knew each other to better reflect a realistic collaborative writing scenario (e.g., writing with a colleague, friend, or classmate). However, five (31\%) mentioned that they may feel less comfortable sharing their prompting activities in scenarios with a power dynamic, such as with a supervisor. Similarly, our study focuses on a synchronous writing context, but one participant mentioned that they may have different opinions when writing asynchronously.

Prior work also suggests that writers may be even more uncomfortable when writing in their non-native language \cite{Strobl2014ForeignWriters}. Although six (37.5\%) participants were not native English speakers, all reported high proficiency (at least ``full professional'' proficiency), so we were not able to study whether this had an effect. 

Finally, 14 participants (87.5\%) reported prior experience using LLMs.
\rev{Although we did not explicitly ask about participant's altitude toward using LLM for writing, 11 (69\%) reported having such experience.} Most (15, 94\%) also reported feeling comfortable overall in collaborative writing scenarios ($\geq$ 5 on a 7-point scale). These results may have contributed towards more positive opinions of sharing more information, especially when using LLMs to support writing.
Repeating our study with different participants, pairs, and a different task would be valuable to better understand how sharing preferences change in other scenarios.

\subsubsection{Other AI Tasks}
Our approach of using comments required participants to select text to create prompts, which may have encouraged them to issue prompts that \emph{edit existing text} instead of \emph{generating new text from scratch}.
It is possible that writers would feel more comfortable sharing text-editing prompts since edits may be associated with more trust and require less verification. However, we were unable to examine the effect of prompt type as we did not ask about comfort for different prompt types.
Similarly, many people use generative AI to receive ideas or feedback \cite{Benharrak2024AIPersonas}, but we did not consider other roles that do not involve writing text. 
It would be valuable to understand how different AI tasks affect willingness to share.

\section{Conclusion}
We explore how writers feel about different levels of prompt information sharing when writing with generative AI in collaborative text editors. Participants wrote persuasive essays in pairs, and our results showed that participants prefer techniques that share more information than those that do not. Participants valued the benefits this had on awareness, even though it was also associated with some drawbacks. As more collaborative writing tools incorporate generative AI capabilities, we believe that it is important for collaborators to have increased transparency over its use, which can be achieved by sharing more information about prompting activities.

\bibliographystyle{lib-acm/ACM-Reference-Format}
\bibliography{main_acm}

\appendix
\makeatother
\clearpage
\renewcommand\thefigure{\thesection.\arabic{figure}}
\renewcommand\thetable{\thesection.\arabic{table}}
\setcounter{figure}{0}
\setcounter{table}{0}

\end{document}